# Modeling the effect of transient violations of the second law of thermodynamics on heat transfer in silicon nanowire

Andrew P. Proudian[1] and Susanta K. Sarkar[1*]
[1]Department of Physics, Colorado School of Mines, USA
*To whom correspondence should be addressed.
E-mail: *ssarkar@mines.edu*

**Abstract**
Violations of the second law of thermodynamics for small systems well below the thermodynamic limit has already been experimentally observed. However, the effects of such violations on experimentally measurable quantities have not been studied in detail. Here we report the effect of transient violations, i.e., spontaneous transfer of heat from cold to hot objects, on heat transfer in silicon nanowire by modeling the violations as a Poisson process. Our modeling and simulations show that it is increasingly difficult to cool down the nanoscale objects to the bath temperature, there is a transition from power law to exponential behavior, and the fluctuations have inverse square law dependence on number of sites.

**Introduction**
Both basic and applied research have seen enormous growth at nanoscale, where the system of interest is either single molecules or small systems of few molecules well below the thermodynamic limit. For example, single-molecule electronics [1-3], microelectronics with node size well below 20 nm [4], artificial molecular motors [5-7], biological molecular motors [8], and driven quantum transport [9] are all related to systems where fluctuations can influence the thermodynamic properties. In addition, most biochemical processes in living systems involve one or few copy numbers of biomolecules. Due to such a broad range of applications, stochastic thermodynamics for small systems have been extensively studied both theoretically and experimentally [10-15]. If $X$ is an extensive quantity proportional to the number of particles $N$ or to the volume of the system $V$, and also if different measurements of $X$ are Poisson distributed, the relative fluctuations of $X$ is given by $\sqrt{\langle (X - \langle X \rangle)^2 \rangle} / \langle X \rangle = 1/\langle X \rangle^{1/2} \sim 1/N^{1/2}$ or $\sim 1/V^{1/2}$. For small systems, the relative fluctuations can be large whether or not the system is in equilibrium (no net transfer of energy or particle) and the fluctuations are Poisson distributed. Due to large relative fluctuations, the probability of transient (not on average) violations of second law, i.e., spontaneous energy transfer from cold to hot objects, becomes high and experimentally measurable [16, 17]. However, the effect of such violations on material properties has not been studied.

In this paper, we have modeled the effect of transient violations of second law on heat transfer in silicon (Si) nanowires (NWs). Si NWs have interesting electronic and optical properties, and are promising for a broad range of applications including electronics [18], lasers [19], photovoltaics [20, 21], and thermoelectrics [22, 23]. In particular, Si NWs have generated interest as thermoelectric materials [22, 23], as the possibility of low thermal conductivity while having high electrical conductivity improves the thermoelectric figure-of-merit, ZT = 2SσT/κ, where σ and κ are respectively the electrical and thermal conductivities, and S is the Seebeck coefficient. The thermal conductivity of Si NWs decreased due to nanoscale dimensions that has been shown both experimentally and computationally [24-31]. We have assumed that the violations are described by a Poisson process, i.e., the violations have a constant probability of happening in time, and have shown that the reduced thermal conductivity could be an effect of transient violations of second law of thermodynamics. Our results motivate a general Poisson process approach to statistical mechanics that can be applied to both non-equilibrium and equilibrium for nanoscale systems.

**Modeling and simulations**

To model heat transfer, we use three experimental physical parameters as inputs to our model: the speed of sound in bulk silicon, c (8433 ms$^{-1}$); the lattice constant of silicon, a (5.431 Å); and the Debye temperature of silicon, $T_D$ (645 K). We assume one-dimensional system with sites connected to their neighbors via a hopping probability given by $P_h = (c/a)dt$, which allows phonons to move through the nanowire. We assume that all phonons have the same frequency (Einstein model) and the Einstein temperature is given by, $T_E = (\pi/6)^{1/3} T_D$. The nanowire is connected at both ends to a heat reservoir at a temperature $T_b$, and phonons can be injected into the

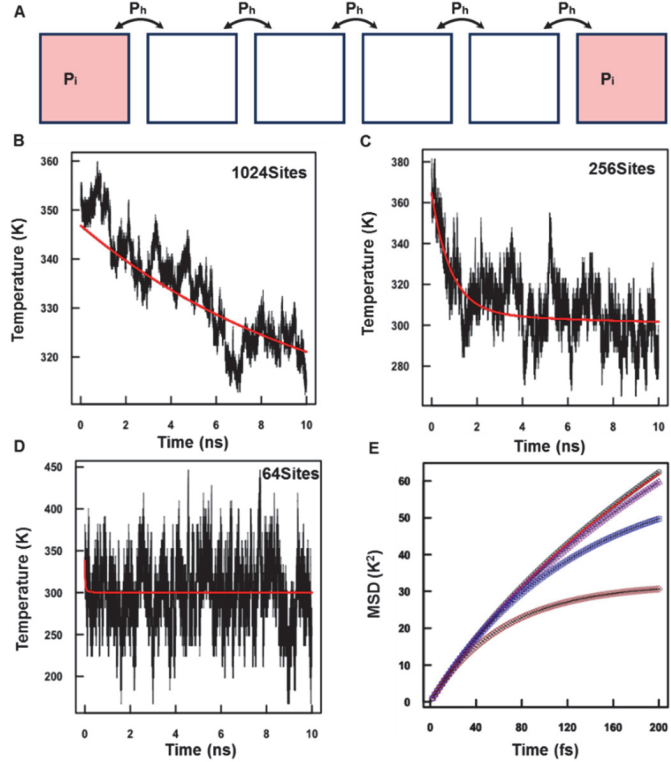

**Figure 1. (A)** A schematic of our model of a silicon nanowire, showing the injection sites and hopping probabilities. Note that the edge injection sites are not included in the nanowire. **(B-D)** The time-series data for the 1024, 256, and 64-site models, respectively. The red curves in Figures 2B-D are the best fits to

$$\overline{\psi}(t) = T_b + \frac{8\Delta T}{\pi^2} \sum_{m=0}^{\infty} \frac{1}{(2m+1)^2} \times \exp\left[-\left(\frac{(2m+1)\pi}{L}\alpha\right)^2 t\right]$$

below 64 sites, the fits did not converge. **(E)** The mean-squared-displacement (MSD) of the first 100 steps for the 1024 (black circles), 64 (red line), 8 (purple triangles), 2 (blue squares), and 1 (brown diamonds) site models, normalized by the corresponding MSD at $\Delta t = 1$.

nanowire from the heat bath with a probability of $P_i = 1/(\exp(T_E/T_b)-1)$. The system is initialized by setting the nanowire's initial temperature to $T_0$ and generating phonons at each site with probability $P_0 = 1/(\exp(T_E/T_0)-1)$. The starting temperature is then calculated from $T_w = T_E/(\log(1+n)-\log(n))$, where $n$ is the average number of phonons per site, not including the edge sites (**Figure 1A**). Once initialized, the model is then time-evolved via a series of steps. First, phonons are injected into the nanowire from the edge sites with probability $P_i$ and each phonon hops with a probability $P_h$. Finally a new system "pseudo temperature" is calculated from the average phonon occupancy $n$ using $T_w = T_E/(\log(1+n)-\log(n))$.

It is important to note that the temperature $T_w$ is simply a transformation of the average number of phonons per site. Temperature is well defined only at equilibrium and we are far from equilibrium in these calculations.

We simulated our model by varying the number of sites from 1 to 1024 in powers of 2. For all of the simulations, we used a time step of $dt = 2$ fs. From the input physical parameters above and equations $P_h = (c/a)dt$ and $P_i = 1/(\exp(T_E/T_b)-1)$, we also used an injection probability of $P_i = 0.215$, and a hopping probability of $P_h = 0.0311$. We considered two initial conditions: 1) the wire at higher temperature ($T_w = 350$ K) and the bath at lower temperature ($T_b = 300$ K); and 2) the wire at lower temperature ($T_w = 300$ K) and the bath at high temperature ($T_b = 350$ K. These choices allowed us to probe both heat flow into and out of the wire for a range of sizes.

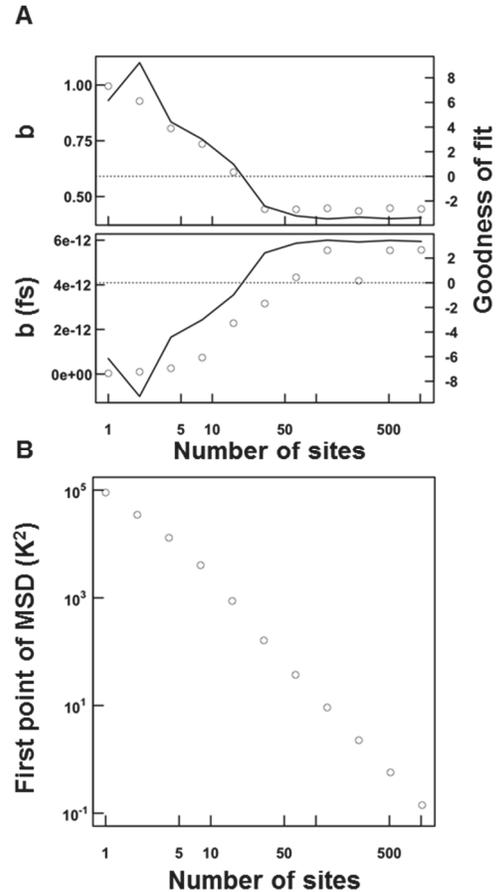

Figure 2. **(A)** MSDs in **Figure 1E** were fitted to a power law equation, $MSD(n\Delta t) = a * t^{1-b}$, and an exponential equation, $MSD = a[1-\exp(-t/b)]$. The upper plot shows the coefficient, $b$, obtained from the power law fits, while the lower plot shows the coefficient, $b$, obtained from the exponential fits. The solid black line is the log-ratio of chi-squared values for two fit models; positive values indicate that the displayed model is a better fit than the alternative, while negative values indicate the alternative is better (the dotted line as zero is a guide for the eye). **(B)** MSD at $\Delta t = 1$ as a function of the number of sites.

**Results and discussion**
The simulated time series of "pseudo temperature" fluctuations for specific number of sites are shown in **Figure 1B-D**. Temperature fluctuations are larger for small number of sites, which results from the coarseness of the average phonon occupancy. However, as the number of sites increases, the temperature follows macroscopic behavior. Using an analytic model (see Supplementary Information), we derive the average wire temperature to be

$$\overline{\psi}(t) = T_b + \frac{8\Delta T}{\pi^2}\sum_{m=0}^{\infty}\frac{1}{(2m+1)^2} \times \exp\left[-\left(\frac{(2m+1)\pi}{L}\alpha\right)^2 t\right]$$

If we fit this equation to our data, with the thermal diffusivity $\alpha$ as the fitting parameter, we get the red curves shown in **Figure 1B-D**. These curves show good agreement with the modeled system temperature on average, but do not display any of the finer behavior evident in the simulated data. In particular, the analytic expression predicts monotonically decreasing temperature as we would expect from the second law while our data clearly show significant deviation from such behavior; indeed there are significant periods where heat moves from the cold bath to the hot wire. This violation has the effect of reducing the thermal diffusivity, as predicted by density functional theory calculations and demonstrated experimentally. Taking the average fit value of 16 runs, we find that for the 1024-site data, we have a thermal diffusivity of $\alpha = 3.3 \pm 1.5 \times 10^{-6}$ m$^2$s$^{-1}$. Comparing this to the data for the 22 nm wire at 300K [25], we find agreement to within our error. To further characterize the heat flow in our model, we look at the mean-squared-displacement (MSD) of simulated temperature fluctuations, which is defined as

$$MSD(n\Delta t) = \frac{1}{M-n} \sum_{i=1}^{M-n} \left[T(t_{i+n}) - T(t_i)\right]^2$$

We look at these curves for some representative cases in **Figure 1E**, where each MSD has been normalized by dividing MSD values with the MSD value at $\Delta t = 1$. The MSD at $\Delta t = 1$ is a measure of fluctuations or noise that is proportional to $1/N^2$ (**Figure 2B**), in contrast to $1/\sqrt{N}$. For all site numbers, MSDs are sub-linear, indicating hindered diffusion of phonons as we expect because the wire is connected to the bath. After normalization, MSDs were fitted to both power law, $MSD(n\Delta t) = a*t^{1-b}$, and exponential, $MSD(n\Delta t) = a[1 - \exp(-t/b)]$. The parameter $a$ obeys a power law with respect to the number of sites, i.e., $a \propto N^\beta$, and an unweighted linear regression gives $\beta = -1.98 \pm 0.06$. Interestingly, **Figure 2B** shows that the MSD at $\Delta t = 1$ also varies with the site number according to a similar power law, and a linear regression gives $\gamma = -2.006 \pm 0.009$. **Figure 2A** shows the fitted values of $b$ for both fits. The power law fits better for smaller number of sites, whereas the exponential fits better for higher number of sits. The goodness of fit has been measured by the log-ratio of chi-squared values for two models; positive values indicate a better fit. In other words, there is a power law to exponential transition above ~50 sites. We identify the parameter $b$ as a measure of the thermal conductivity.

**Conclusions**
In summary, we have developed a Poisson process approach to model the effects of transient violation of second law of thermodynamics in Si NWs. We have taken a probabilistic approach by assuming that the violations have a constant probability of happening at each time step. Our approach is applicable for both equilibrium and non-equilibrium conditions. The results show that it is increasingly difficult for NWs to attain the bath temperature. For nanoscale, unlike the macroscale, the sample temperature may not be the same as the bath temperature during the experiments. Fluctuations in the system follows an inverse square law behavior in contrast to the normal inverse square root law and there is a transition from power law to exponential as the number of sites increases. Our Poisson process approach to statistics can be easily extended to other systems as well.

**Acknowledgments**
S.K.S. has been supported by the Professional Development Fund at the Colorado School of Mines. A.P.P. acknowledge the support of his advisor, Jeramy Zimmerman.

**Supplementary information**
**1. Analytical model of heat transfer in one dimension**
We begin with the heat equation
$$\frac{\partial \psi}{\partial t} = \alpha \frac{\partial^2 \psi}{\partial x^2}$$
where $\alpha$ is the thermal diffusivity. Assuming a solution of the form $\psi(x,t) = \chi(x)(\tau)$, we get
$$\frac{1}{\tau(t)}\frac{d\tau}{dt} = \frac{\alpha}{\chi(x)}\frac{d^2\chi}{dx^2}$$
Since two sides depend on two different variables, they must be constant. Therefore, we have two equations
$$\frac{d\tau}{dt} = -\lambda\tau$$
$$\frac{d\tau}{dt} = -\lambda\tau \frac{d^2\chi}{dx^2} = -\frac{\lambda}{\alpha}\chi$$
whose solutions for $\lambda \geq 0$ are
$$\tau(t) = Ae^{-\lambda t} + C_0$$
$$\chi(x) = A\cos\left(\sqrt{\frac{\lambda}{\alpha}}x\right) + B\sin\left(\sqrt{\frac{\lambda}{\alpha}}x\right) + C_0'x + C_0$$
Finally, we arrive at the solution
$$\psi(x,t) = [A\cos(\omega x) + B\sin(\omega x)]e^{-\alpha\omega^2 t} + C_0'x + C_0$$
where $\omega \equiv \sqrt{\lambda/\alpha}$. We now apply the particular conditions of our model setup. We begin with the boundary conditions shown in **Figure 1A**,
$$\psi(-L/2, t) = T_b$$
$$\psi(+L/2, t) = T_b$$
$$\psi(x, 0) = T_0$$
Such conditions require that
$$C_0' = B = 0$$
$$C_0 = T_b$$
leading to
$$\psi(x, 0) = \sum_{n=1}^{\infty} a_n \cos\left(\frac{n\pi x}{L}\right) + T_b = T_0$$
$$\Rightarrow \Delta T = T_0 - T_b = \sum_{n=1}^{\infty} a_n \cos\left(\frac{n\pi x}{L}\right)$$
Inverting $\Delta T$ for $a_n$ leads to
$$a_n = \frac{2}{L}\int_{-L/2}^{L/2} dx \Delta T \cos\left(\frac{n\pi x}{L}\right) = \frac{(-1)^m 4\Delta T}{(2m+1)\pi}$$
where $n = 2m+1$. Thus, our particular solution is

$$\psi(x,t) = T_b + \frac{4\Delta T}{\pi} \sum_{m=0}^{\infty} \frac{(-1)^m}{2m+1} \cos\left(\frac{(2m+1)\pi}{L}x\right) \times \exp\left[-\alpha\left(\frac{(2m+1)\pi}{L}\right)^2 t\right]$$

This solution, however, is too detailed to apply to our model system in Figure 1A, as we only measure the mean temperature of our wire over time. Therefore, we must take the mean value of this function over its entire length,

$$\overline{\psi}(t) \equiv \frac{1}{L}\int_{-L/2}^{L/2} dx\, \psi(x,t)$$

which, upon evaluating the integral, gives

$$\overline{\psi}(t) = T_b + \frac{8\Delta T}{\pi^2} \sum_{m=0}^{\infty} \frac{1}{(2m+1)^2} \times \exp\left[-\left(\frac{(2m+1)\pi}{L}\alpha\right)^2 t\right]$$

## 2. Exact calculation of thermodynamic quantities

We begin by counting the number of ways $q$ identical particles can be arranged on $N$ sites

$$\Omega = \frac{(q+N-1)!}{q!(N-1)!}$$

which can be rewritten as an analytic expression using the gamma function as

$$\Omega = \frac{\Gamma(q+N)}{\Gamma(q-1)\Gamma(N)} = \frac{q\Gamma(q+N)}{\Gamma(q)\Gamma(N)} = \frac{q}{B(q,N)}$$

where we have used the property of the gamma function $\Gamma(z+1) = z\Gamma(z)$ and the definition of beta function

$$B(a,b) \equiv \frac{\Gamma(a)\Gamma(b)}{\Gamma(a+b)}$$

As the entropy is defined by $S \equiv k\ln(\Omega)$, we have $S = k\left[\ln q - \ln B(q,N)\right]$

### 2.1. Temperature

For our system of harmonic oscillators, the energy (ignoring the zero-point energy) is $U = q\varepsilon$, where $\varepsilon \equiv \hbar\omega$. Thus, our system temperature may be calculated from

Using the equation $S = k\left[\ln q - \ln B(q,N)\right]$, we can calculate

$$\frac{\partial S}{\partial q} = k\left(\frac{1}{q} - \frac{1}{B(q,N)}\frac{\partial B(q,N)}{\partial q}\right)$$

$$= k\left(\frac{1}{q} - \frac{B(q,N)[\psi_0(q) - \psi_0(q+N)]}{B(q,N)}\right)$$

$$= k\left(\frac{1}{q} + \psi_0(q+N) - \psi_0(q)\right)$$

where $\psi_0$ is the digamma function. As both $q$ and $N$ are integers, we can use the relation

$$\psi_0(n) = -\gamma + \sum_{m=1}^{n-1}\frac{1}{m}$$

and write

$$\frac{\partial S}{\partial q} = k\left(\frac{1}{q} + \sum_{m=1}^{q+N-1}\frac{1}{m} - \sum_{m=1}^{q-1}\frac{1}{m}\right)$$

$$= k\left(\frac{1}{q} + \sum_{m=1}^{q+N-1}\frac{1}{m}\right)$$

$$= k\left(\frac{1}{q} + \sum_{m=1}^{N-1}\frac{1}{m+q}\right)$$

$$= \sum_{m=0}^{N-1}\frac{k}{m+q}$$

Therefore,

$$\frac{1}{T} = \frac{1}{\varepsilon}\frac{\partial S}{\partial q} = \frac{k}{\varepsilon}\sum_{m=0}^{N-1}\frac{1}{m+q}$$

By using $\theta_E = \varepsilon/k$, we have

$$\frac{1}{T} = \frac{1}{\theta_E}\sum_{m=0}^{N-1}\frac{1}{m+q}$$

**2.2. Heat capacity**

The specific heat is given by

$$c_V = \left(\frac{\partial U}{\partial T}\right)_V$$

which we can rewrite as

$$\frac{1}{c_V} = \frac{\partial T}{\partial U} = \frac{1}{\varepsilon}\frac{\partial T}{\partial q}$$

Using the equation

$$\frac{1}{T} = \frac{\partial S}{\partial U} = \frac{1}{\varepsilon}\frac{\partial S}{\partial q}$$

we have

$$\frac{\partial T}{\partial q} = \theta_E \frac{\displaystyle\sum_{n=0}^{N-1}\frac{1}{(q+n)^2}}{\left(\displaystyle\sum_{n=0}^{N-1}\frac{1}{q+n}\right)^2}$$

Therefore, the specific heat is

$$c_V = k\frac{\left(\displaystyle\sum_{n=0}^{N-1}\frac{1}{q+n}\right)^2}{\displaystyle\sum_{n=0}^{N-1}\frac{1}{(q+n)^2}}$$

where we have the definition $\theta_E \equiv \varepsilon/k$.

# 3. Variance of temperature fluctuations

Table I

| Number of sites (N) | $2^0, 2^1, ..., 2^{20}$ |
|---|---|
| Temperature of bath ($T_b$) | 300 K |
| Einstein temperature ($T_e$) | ~519 K |
| Phonon occupancy of bath ($n_b$) | ~0.214 |
| Mean number of phonons ($\mu$) | $n_b N$ |
| Standard deviation of phonons ($\sigma$) | 2 |

To check the temperature calculations, we studied the effects of noise as a function of number of sites. We generated Gaussian distributed phonon fluctuations for different site number, $N$. **Table I** shows the parameters that we have used. The mean number of phonons change with $N$ to keep the temperature constant, but have the same phonon noise for all $N$. This resembles our simulated heat transfer data in **Figure 1**. With simulated phonon number data, we calculated temperatures using two definitions, i.e., the exact temperature as shown in the paper and the Einstein temperature valid for large systems, and calculated MSDs at lagtime, $\Delta t = 1$. For the simulated phonon number data, MSD($\Delta t = 1$) is equal to $2\sigma^2$ which is $8.2 \pm 0.5 \sim 2*2^2 = 8$ for all $N$. However, the two calculated temperature sets both show $1/N^2$ dependence. Unweighted linear fits to the log-transformed temperature data give the exponent as $-1.977 \pm 0.012$ (exact) and $-1.996 \pm 0.005$ (Einstein).

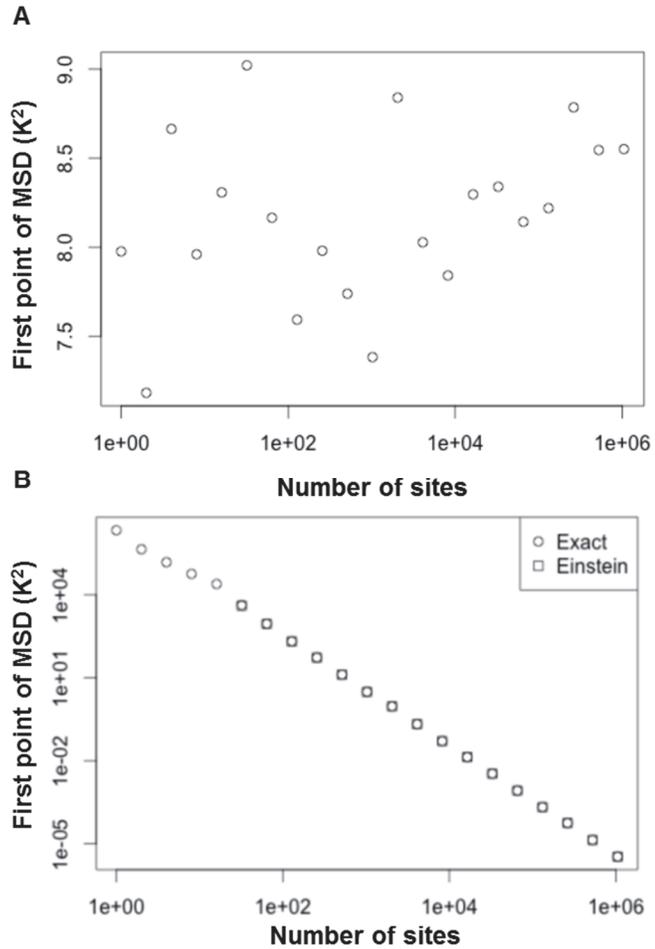

**Figure S1. (A)** The first-order MSD for the phonon number data; the first-order MSD is constant with $N$ and is equal to $2\sigma^2$ (i.e. $8:2 \pm 0.5 \sim 2 * 2^2 = 8$). **(B)** MSD at $\Delta t = 1$ as a function of the number of sites. The first-order MSD for the transformed temperature data using both Einstein and exact transformations; both show the $1/N^2$ dependence.